# Inactivated COVID-19 Vaccination did not affect In vitro fertilization (IVF) / Intra-Cytoplasmic Sperm Injection (ICSI) cycle outcomes

Qi Wan[a,b,c#], Ying-Ling Yao[d,e#], Xing-Yu Lv[c#], Li-Hong Geng[c], Yue Wang[e], Enoch Appiah Adu-Gyamfi[g], Xue-Jiao Wang[c], Yue Qian[e], Juan Yang[c], Ming-Xing Chen[d,e], Zhao-Hui Zhong[e], Yuan Li[c]*, and Yu-Bin Ding[d,e,f]*

[a] Department of Obstetrics and Gynecology, West China Second Hospital, Sichuan University, Chengdu, 610041, China

[b] Key Laboratory of Birth Defects and Related Diseases of Women and Children, Ministry of Education, Sichuan University, Chengdu, 610041, China

[c] Sichuan Jinxin Xinan Women & Children's Hospital, Chengdu, 610011 Sichuan, China

[d] Department of Obstetrics and Gynecology, Women and Children's Hospital of Chongqing Medical University, Chongqing, 401147, China

[e] Joint International Research Laboratory of Reproduction and Development of the Ministry of Education of China, School of Public Health, Chongqing Medical University, Chongqing, 400016, China

[f] Department of Pharmacology, Academician Workstation, Changsha Medical University, Changsha, 410219, China

[g] Department of Biomedical Sciences, State University of New York at Albany, New York, 12208, USA

*Corresponding authors:

Yu-Bin Ding

Department of Obstetrics and Gynecology, Women and Children's Hospital of Chongqing Medical University, Chongqing, China

No.120 Longshan Road, Yubei District, Chongqing, 401147, China Email: dingyb@cqmu.edu.cn

Yuan Li

The Reproductive Center

Sichuan Jinxin Xinan Women & Children's Hospital

No.66 and No.88, Bi Sheng Road, Jinjiang District, Chengdu, 610011, China

E-mail: llyydoctor@163.com

Abstract

Background: The objective of this study is to evaluate the impact of COVID-19 inactivated vaccine administration on the outcomes of in vitro fertilization (IVF) and intracytoplasmic sperm injection (ICSI) cycles in infertile couples in China.

Methods: We collected data from the CYART prospective cohort, which included couples undergoing IVF treatment from January 2021 to September 2022 at Sichuan Jinxin Xinan Women & Children's Hospital. Based on whether they received vaccination before ovarian stimulation, the couples were divided into the vaccination group and the non-vaccination group. We compared the laboratory parameters and pregnancy outcomes between the two groups.

Findings: After performing propensity score matching (PSM), the analysis demonstrated similar clinical pregnancy rates, biochemical pregnancy and ongoing pregnancy rates between vaccinated and unvaccinated women. No significant disparities were found in terms of embryo development and laboratory parameters among the groups. Moreover, male vaccination had no impact on patient performance or pregnancy outcomes in assisted reproductive technology treatments. Additionally, there were no significant differences observed in the effects of vaccination on embryo development and pregnancy outcomes among couples undergoing ART.

Interpretation: The findings suggest that COVID-19 vaccination did not have a significant effect on patients undergoing IVF/ICSI with fresh embryo transfer. Therefore, it is recommended that couples should receive COVID-19 vaccination as scheduled to help mitigate the COVID-19 pandemic.

Funding: This work was supported by the Scientific and Technological Research Program of Chongqing Municipal Education Commission (grant no. KJZD-K202200408) and the Science and Technology Department of Sichuan Province (grant no. 21PJ166).

Introduction

In the context of the COVID-19 pandemic, the administration of COVID-19 vaccines has become a crucial strategy to mitigate the spread of the virus and its associated impact on public health. As the vaccination efforts continue, it is essential to assess the effects of these vaccines on various aspects of health, including their impact on fertility and assisted reproductive technology outcomes.

With the rapid spread of COVID-19, there has been a pressing need to understand the safety and effectiveness of COVID-19 vaccines. However, there is limited information available on the specific effects of these vaccines on fertility and assisted reproductive technology outcomes. This knowledge gap has led to concerns among individuals who are planning to undergo fertility treatments, particularly IVF and ICSI, as well as those considering the timing of conception and vaccine administration.

To date, studies have offered preliminary evidence regarding the safety of COVID-19 vaccines in relation to fertility. Meta-analyses consistently indicate that COVID-19 vaccines have no significant impact on fertility in both females and males[1-6]. These studies consistently report no notable differences in the number of obtained, produced, and matured oocytes among individuals undergoing assisted reproductive technology, regardless of their vaccination status, COVID-19 infection, or lack of vaccination[7]. mRNA SARS-CoV-2 vaccines do not appear to affect the treatment outcomes or ovarian reserve in subsequent IVF cycles[8]. Furthermore, vaccination does not exert any detrimental effects on male semen parameters[9-11]. Meta-analysis results suggest that mRNA vaccines are highly effective against SARS-CoV-2 during pregnancy. The incidence of adverse pregnancy outcomes did not increase among pregnant women who received the vaccine compared to those who were unvaccinated. In fact, the occurrence of stillbirth and preterm birth was lower among vaccinated pregnant women. Importantly, based on data from large national registries and reports of inadvertent exposure during early pregnancy in randomized controlled trials, the risk of miscarriage did not increase after receiving the COVID-19 vaccine[12,13]. These findings provide a certain level of assurance. However, most studies have primarily examined the impact of mRNA vaccines, and evidence on the safety of inactivated vaccines for IVF outcomes is limited. Additionally, data on the safety of COVID-19 vaccines during early pregnancy is limited. Therefore, further comprehensive investigations are crucial for better understanding the potential impact of administering inactivated COVID-19 vaccines on IVF and ICSI outcomes.

Currently, most studies have focused on investigating the pregnancy outcomes of women who received the vaccine, while lacking information on male partners' vaccination status. This study incorporates information on male vaccination to analyze the impact of COVID-19 inactivated vaccine administration on IVF and ICSI outcomes among infertile couples. Understanding the effects of COVID-19 vaccination on fertility and assisted reproductive technology outcomes is crucial for healthcare providers and individuals undergoing fertility treatments. This knowledge will help make informed decisions and provide appropriate guidance.

## Materials and methods

### Study design and participants

We collected data from the CYART prospective cohort, which consisted of couples undergoing IVF treatment January, 2021 to September, 2022 in Sichuan Jinxin Xinan Women & Children's Hospital. This cohort study was approved by the Ethics Committee of Sichuan Jinxin Xinan Women & Children's Hospital (2021-014).

Exclusion criteria were a history of COVID-19 infection, based on their self-reports and regulations from China Zero Policy, none of the participants had a history of being infected with the Covid-19 virus; patients with donor sperm or oocyte cycle.

Questionnaires were used to obtain the vaccination information about each study participant. The questionnaires were distributed through the WeChat app, and the study participants filled them out voluntarily. The contents encompassed information regarding the timing, dosage, and manufacturers of the vaccine. Participants can obtain accurate immunization records through personal accounts on various mobile applications in China, including Sichuan Tianfu Tong, Alipay, and WeChat. Electronic records of individuals were established in local health departments, where individuals reported their vaccination status. This information was collected from personal accounts on the social networking platform WeChat and the official application Sichuan Tianfu Tong, which is initiated, supported, and operated by the local government department. All patients received inactivated Covid-19 vaccines from three manufacturers: Beijing Institute of Biological Products, Wuhan Institute of Biological Products and Beijing Sinovac Biotech Co., Ltd. Each vaccination program involved three serial doses of the vaccine. The interval between the first and second doses was at least 14 days, while the interval between the second and third doses was at least 6 months.

The vaccine group consisted of couples who received vaccination before ovarian stimulation, whereas the control group comprised couples who did not receive vaccination prior to ovarian stimulation.

### Ovarian stimulation protocols

In accordance with the patient's age, body mass index (BMI), antral follicle number (AFC), and anti-Mullerian hormone levels (AMH), recombinant follicle stimulating hormone (rFSH, Gonal-F, Merck Serono S. A., Switzerland) 100-300 IU/d administration was performed from the 2nd to the 4th day of the menstrual cycle or on the 28th to the 35th day after gonadotropin-releasing hormone (GnRH) agonist administration. The gonadotropin (Gn) dosage was adjusted as the follicles developed.

A daily dose of 0.25 mg Gonadotropin-releasing hormone antagonist (GnRH-ant) was administered starting from the 6th day of rFSH stimulation until the administration of human chorionic gonadotropin (hCG) or when the dominant follicle reached a diameter of ≥ 12-14 mm. The GnRH-ant used in this study was either Ganerik acetate (Merck Serono, Switzerland) or Ganirelix (Vetter Pharma-Fertigung GmbH & Co.KG). The induction of ovulation was performed by administering women with 250 μg of recombinant human chorionic gonadotropin (rhCG, Merck Schlano, Germany) or with 0.2 mg of Decapeptyl (Ferring GmbH, Germany) either alone or in combination with 2000 IU of urinary hCG (LiZhu, China). This was performed during the period when two to three ovarian follicles were, at least, 17-18 mm in diameter. Oocyte retrieval was performed 36-38 h after the induction of ovulation.

Embryo Transfer and Luteal Support

On the 3rd to 5th day after fertilization, 1 to 2 of grade I-II high-quality embryos were selectively transferred. The luteal phase support was started on the day when the oocytes were retrieved with 90 mg of vaginal progesterone gel (Merck Schlano, Germany). A 10 mg of dydrogesterone (Dupbaston, Dutch) and 4 mg of estradiol valerate（Bayer，Germany）was taken twice each day.

Outcomes

The primary outcome measure was the clinical pregnancy, which was defined as the presence of at least one gestational sac observed in the uterine cavity through ultrasound examination conducted four weeks after embryo transfer. Secondary outcome measures included the number of oocytes retrieved, the number of blastocysts formed, implantation rate, biochemical pregnancy, and ongoing pregnancy. Biochemical pregnancy was defined as having serum β-human chorionic gonadotropin (β-hCG) levels greater than 25 IU/L at 14 days after embryo transfer. Ongoing pregnancy referred to a clinical pregnancy that persisted for at least 12 weeks.

Statistical analysis

The following aspects of the vaccinated versus unvaccinated groups were compared: baseline characteristics, ovarian stimulation parameters, embryo development, and pregnancy rates. Statistical analyses were performed using R (version 4.0.3). Quantitative variables with normal distribution and homogenous variance are expressed as mean ± standard deviation (SD), with the means compared using Students' t-test. Quantitative variables with abnormal distribution or heterogeneous variance are expressed as median (IQR). The medians were compared by using Mann-Whitney U-test. Differences in the rates were compared by using Chi-squared test. When the expected count was <5 or the total sample size was <40, Fisher's exact test was performed to compare the differences in the rates. Statistical significance was set at P < 0.05.

Propensity score matching (PSM) was applied to screen a group of patients, such that the baseline parameters of the vaccinated group were quite similar to those of the unvaccinated group. The propensity score was calculated using a multiple logistic regression model. We performed propensity score matching using the MatchIt package, with a caliper width for propensity score matching set at 0.2 times the standard deviation (SD). The matching ratio was 1:2, and we employed nearest neighbor matching. The standard deviation of the independent variables before and after propensity score matching was calculated. An absolute value of SD<10% was considered indicative of balance.

Sensitivity Analyses

Conducting sensitivity analysis using a multiple logistic regression model to assess the impact of COVID-19 vaccination on pregnancy outcomes in females or males.

## Results

### Vaccination of infertile couples

A total of 9,044 couples undergoing fresh embryo transfer were recruited between January 1, 2021, and September 30, 2022. Excluding patients without vaccine information, a total of 722 couples were included. A total of 511 female received the vaccine before controlled ovarian stimulation (COS). These 511 women were categorized into the vaccination group, while the remaining 211 female were placed in the unvaccinated group. Additionally, 26.5% (191/722) of male partners were also vaccinated. A total of 171 couples were vaccinated, while 191 couples remained unvaccinated (Figure 1).

### IVF outcomes stratified by female vaccination status

Table 1 presents the baseline characteristics grouped by female vaccination status before (left) and after (right) matching. Prior to propensity score matching (PSM), the two groups exhibited significant differences in infertility duration, infertility diseases, and ovarian stimulation protocol. However, after PSM, the baseline characteristics of vaccinated and unvaccinated women were well balanced. Following the propensity score matching, the vaccinated group consisted of 355 women, while the unvaccinated group had 211 women. On the day of hCG, there were similar numbers of dominant follicles (≥14mm or 17mm) as well as comparable counts of retrieved oocytes, mature eggs, fertilized eggs, MII eggs, 2pn fertilized eggs, and cleaved embryos between the vaccinated and unvaccinated groups. Additionally, the ratios of high-quality day 3 embryos, high-quality blastocysts, and blastocyst formation were similar in both groups (Table 2).

The results showed no differences in clinical pregnancy, biochemical pregnancy, and ongoing pregnancy between the vaccinated and unvaccinated groups before and after propensity score matching (Table 2).

Sensitivity analyses reported largely consistent results in all aforementioned outcomes (Figure 3).

IVF outcomes stratified by male vaccination status

An analysis was performed on the male vaccination status groups, revealing no differences in baseline characteristics (Table 3) and laboratory parameters (Table 4) between the vaccinated and unvaccinated groups. The vaccination status of couples undergoing assisted reproductive technology did not significantly impact the ongoing pregnancy rate, biochemical pregnancy rate, and clinical pregnancy rate (Table 4).

Sensitivity analyses reported largely consistent results in all aforementioned outcomes (Figure 4).

IVF outcomes stratified by couples' vaccination status

Baseline demographics among the controlled ovarian hyperstimulation cohort are shown in Table 5. Compared to the vaccination group, the unvaccinated group had higher BMI and infertility duration. There were differences between the two groups in terms of fertilization type and ovarian stimulation protocols. Furthermore, the vaccination status of couples undergoing assisted reproductive technology does not significantly affect embryo development (e.g., follicle count, oocytes retrieved, mature oocytes, fertilized eggs, MII oocytes, 2PN oocytes, high-quality blastocysts, and blastocyst count) or pregnancy outcomes, including ongoing pregnancy rate, clinical pregnancy rate, and biochemical pregnancy rate (Figure 2).

Discussion

The safety of COVID-19 vaccines in women undergoing IVF/ICSI treatment remains uncertain. Understanding the effects of COVID-19 vaccine administration on fertility potential and ART outcomes is crucial for healthcare providers and couples seeking fertility treatment. Our data suggest that couples undergoing assisted reproductive technology who receive inactivated vaccines do not experience any impact on embryo development and pregnancy outcomes. These findings provide valuable insights into the potential effects of inactivated COVID-19 vaccine administration on assisted reproduction outcomes for this specific population.

In recent years, inactivated vaccines against COVID-19 have been widely available worldwide

[14]. Numerous reports suggest that COVID-19 vaccination does not pose any risks to women planning for pregnancy, pregnant women, or those who are breastfeeding [15-21]. Recently, several studies have reported that the administration of COVID-19 mRNA vaccines did not demonstrate any adverse effects on ovarian stimulation or early pregnancy outcomes in the context of in vitro fertilization[22,23]. However, most of the existing evidence focuses on mRNA vaccines, and there is limited safety data available on inactivated vaccines in the context of assisted reproduction. Unlike live attenuated vaccines, inactivated vaccines have not shown any notable adverse effects on maternal or fetal health and are generally considered safe for pregnant and preconceptional women [24,25]. Our results provide preliminary evidence confirming the safety of inactivated COVID-19 vaccines for couples planning to undergo IVF.

The emergence and rapid spread of the COVID-19 in 2020 have had detrimental effects on pregnant women and their newborns. There are reports indicating an increased risk of pre-eclampsia, preterm birth, gestational diabetes, low birth weight, and stillbirth among pregnant women infected with the novel coronavirus[26]. COVID-19 infection may also lead to significantly lower scores in overall motor skills, fine motor skills, and personal-social domains among infants[26,27]. While SARS-CoV-2 has not been directly isolated from the human endometrium, there remains a potential risk of infection in the uterine endometrium, particularly during the crucial implantation window[28,29]. The presence of virus-infectivity-related genes ACE2 and TMPRSS4 in the human endometrium suggests that SARS-CoV-2 could potentially invade and infect cells, leading to tissue damage. The expression levels of these virus-infectivity-related genes increase from the proliferative phase to the secretory phase, indicating a potential vulnerability to SARS-CoV-2 infection during the embryo implantation period [30,31]. Furthermore, COVID-19 infection may result in a decline in ovarian function, alterations in endometrial receptivity, and reduced semen quality in males, with an estimated recovery time of three months[32-36]. Given the absence of a specific approved cure for the disease in pregnant women, widespread vaccination serves as a crucial approach adopted by countries to mitigate the pandemic's impact on this particular population. Therefore, considering the infectious risk associated with SARS-CoV-2, it is necessary for couples planning embryo transfer to receive the COVID-19 vaccine before commencing assisted reproductive technology cycles in order to prevent virus infection.

Several studies have examined the relationship between COVID-19 vaccine administration and male fertility. In two studies involving couples undergoing fertility treatment and one study involving the general population, no significant differences were observed in semen volume, sperm concentration, or vitality before and after COVID-19 vaccine administration[9,37,38]. However, due to the variability and susceptibility of semen analysis, we conducted an analysis of IVF outcomes in our participants, providing direct evidence regarding the impact of inactivated vaccines on gamete viability and embryonic development potential. Our results suggest that there were no significant differences in laboratory and clinical outcomes of IVF, indicating that male vaccination with inactivated vaccines is unlikely to have adverse effects on IVF outcomes. However, due to the relatively small number of male participants who received the COVID-19 vaccine, it is crucial to interpret this conclusion with caution. Future research with larger sample sizes is needed to validate these findings and further investigate the potential impact of male vaccine administration on reproductive outcomes.

Of course, this study has several limitations. It is a single-center study and not a multi-center study, which limits the generalizability of the results. Multi-center studies involving diverse populations are needed to obtain a more comprehensive understanding of the relationship between COVID-19 vaccine administration and IVF outcomes. Furthermore, longer follow-up periods are crucial for observing pregnancy outcomes beyond the early stages of pregnancy, which were relatively short in our study. Additionally, the different types of COVID-19 vaccines and vaccination strategies used in various countries and regions may influence the study results.

In conclusion, the evidence provided by this study suggests that COVID-19 inactivated vaccine administration does not have an adverse impact on IVF outcomes. Pregnancy is not a contraindication for COVID-19 vaccines, especially inactivated vaccines. In situations where there is a risk of contracting the novel coronavirus, it is recommended to receive the COVID-19 vaccine as per the established guidelines. This information is crucial for couples planning to conceive, especially those undergoing ART, as it alleviates concerns about the potential impact of vaccine administration on pregnancy and ART results. The findings contribute to understanding the safety of COVID-19 vaccines for fertility and IVF outcomes, offering valuable insights for healthcare professionals, couples planning to conceive, and policymakers involved in vaccination programs and reproductive health.

## Conclusions

Our study showed no difference in IVF data between vaccinated and unvaccinated couples, and that vaccination did not adversely affect IFV/ICSI. Couples attempting ART should not postpone their COVID-19 vaccination because of their ovarian stimulation and embryo transfer schedules.

## Author contributions

Qi Wan, Ying-Ling Yao and Xing-Yu Lv made equal contributions to this work.

Qi Wan drafted the original manuscript, while Ying-Ling Yao and Xing-Yu Lv conducted data analysis. Yu-Bin Ding and Yuan Li participated in the project design and manuscript revision. Enoch Appiah Adu-Gyamfi provided language revision for the manuscript. Zhao-Hui Zhong, Yue Qian, Yue Wang, and Ming-Xing Chen were involved in the design of the questionnaire, data collection, and manuscript revision. Li-Hong Geng, Juan Yang, and Xue-Jiao Wang provided clinical consultation and assisted in data collection.

## Conflict of interest

The authors report no conflict of interest.

## Funding

This work was supported by the Scientific and Technological Research Program of Chongqing Municipal Education Commission (grant no. KJZD-K202200408) and the Science and Technology Department of Sichuan Province (grant no. 21PJ166).

## Data sharing statement

Request for access to the data should be made to the corresponding author at dingyb@cqmu.edu.cn. Data could be made available provided the applicant has appropriate ethics and author approval.

Table 1 Baseline characteristics grouped by the female vaccination status

|  | Before PSM | | | After PSM | | |
|---|---|---|---|---|---|---|
|  | Vaccinated | Unvaccinated | P-value | Vaccinated | Unvaccinated | P-value |
| n | 511 | 211 |  | 355 | 211 |  |
| Age (years) | 31.0 (28.0,34.0) | 31.0 (28.0,34.0) | 0.713 | 31.0 (28.0,34.0) | 31.0 (28.0,34.0) | 0.810 |
| Infertility duration (years) | 3.0(2.0,4.0) | 2.0(1.0,4.0) | 0.036 | 2.0 (1.0,4.0) | 3.0 (2.0,4.0) | 0.081 |
| AMH (ng/ml) | 3.0 (2.8,4.4) | 2.8 (1.9,4.0) | 0.077 | 2.8 (1.9,4.0) | 3.0 (2.1,4.4) | 0.049 |
| Total AFC | 15.0 (11.0,21.0) | 15.0 (11.0,19.0) | 0.643 | 15.0 (11.0,19.0) | 15.0 (11.0,21.0) | 0.710 |
| BMI (kg/m2) | 21.9 (20.0,24.4) | 21.5 (20.0,23.4) | 0.174 | 21.5 (20.0,23.4) | 21.8 (20.0,24.2) | 0.447 |
| Infertility diseases, n (%) |  |  | 0.032 |  |  | 0.881 |
| Tubal factor | 290(56.8) | 127(60.2) |  | 127 (60.2) | 213 (60.0) |  |
| Ovarian factor | 36(7.0) | 14(6.6) |  | 14 (6.6) | 22 (6.2) |  |
| Male factor | 101(19.8) | 24(11.4) |  | 24 (11.4) | 48 (13.5) |  |
| Unexplained | 92(16.4) | 38(21.9) |  | 46 (21.8) | 72 (20.3) |  |
| Type of infertility, n (%) |  |  | 0.744 |  |  | 0.900 |
| Primary | 252(71.4) | 101(28.6) |  | 110 (52.2) | 187 (52.7) |  |
| Secondary | 259(70.2) | 110(29.8) |  | 101 (47.9) | 168 (47.3) |  |
| Parity | 1(0,2) | 1(0,2) | 0.405 | 1 (0,2) | 1 (0,2) | 0.619 |
| Gravidity | 0(0,0) | 0(0,0) | 0.984 | 0(0,0) | 0(0,0) | 0.738 |
| Basal FSH | 7.6(6.5,8.6) | 7.4(6.4,8.8) | 0.927 | 7.4 (6.3,8.8) | 7.6 (6.4,8.4) | 0.858 |
| Basal E2 | 33.0 (26.0,41.0) | 33.0 (27.0,41.0) | 0.760 | 33.0 (26.0,41.0) | 33.0 (26.0,42.0) | 0.415 |
| Basal P | 0.5(0.4,0.8) | 0.5(0.3,0.8) | 0.303 | 0.50(0.3,0.7) | 0.51(0.3,0.8) | 0.375 |
| Basal PRL | 256.0(188.3,352.9) | 261.4(186.7,356.6) | 0.637 | 262.8(187.2,369.6) | 251.4(179.6,349.1) | 0.502 |
| Basal LH | 3.7(2.7,4.9) | 3.7(2.7,5.2) | 0.934 | 3.7(2.6,5.0) | 3.7 (2.7,4.9) | 0.958 |
| LH level on trigger day (mIU/mL) | 1.1(0.7,2.3) | 1.1(0.7,1.8) | 0.419 | 1.1(0.7,1.8) | 1.1 (0.7,2.1) | 0.651 |
| P level on trigger day (ng/mL) | 0.7(0.5,1.0) | 0.7(0.5,0.9) | 0.608 | 0.8 (0.4,0.9) | 0.7 (0.5,0.9) | 0.502 |
| E2 level on trigger day (pg/mL) | 1821.0(1349.3,2584.2) | 1855.0(1309.0,2605.0) | 0.973 | 1842.0(1314.0,2561.5) | 1859.0(1376.0,2590.5) | 0.528 |
| Ovarian stimulation protocols, n (%) | | | 0.032 | | | 0.896 |
| Antagonist | 263(51.5) | 89(42.2) | | 178 (84.4) | 298 (83.9) | |
| Agonist | 248(48.5) | 122(57.8) | | 33 (15.6) | 57 (16.1) | |
| Fertilization type, n (%) | | | 0.120 | | | 0.131 |
| IVF | 404(79.1) | 178(84.4) | | 122 (57.8) | 182 (51.3) | |
| ICSI | 107(20.9) | 33(15.6) | | 89 (42.2) | 173 (48.7) | |
| Endometrial thickness(mm) on the day before ET | 11.0 (9.5,12.0) | 11.0 (9.5,12.5) | 0.226 | 11.0 (9.8,12.5) | 11.0 (9.5,12.0) | 0.322 |
| Total gonadotropin dose (IU) | 2025.0(1650.0,2400.0) | 2025.0(1668.0,2475.0) | 0.374 | 2025.0(1687.5,2475.0) | 2025.0(1650.0,2400.0) | 0.434 |
| Stimulation duration (days) | 10(9,12) | 11(9,12) | 0.074 | 11 (9,12) | 10 (9,12) | 0.190 |
| No. of embryos transferred, n (%) | | | 0.499 | | | 0.594 |
| 1 | 185(36.2) | 82(38.9) | | 82 (38.9) | 130 (36.6) | |
| 2 | 326(63.8) | 129(61.10 | | 129 (61.1) | 225 (63.4) | |
| No. of high-quality embryos for transfer, n (%) | | | 0.467 | | | 0.239 |
| 0 | 89(17.4) | 45(21.3) | | 45 (21.3) | 56 (15.8) | |
| 1 | 269(52.6) | 105(49.8) | | 105 (49.8) | 193 (54.4) | |
| 2 | 153(29.9) | 61(28.9) | | 61 (28.9) | 106 (29.9) | |
| Days of embryos transferred, n (%) | | | 0.683 | | | 0.502 |
| D3 | 263(51.5) | 105(49.8) | | 105 (49.8) | 187 (52.7) | 0.810 |
| D5 | 248(48.5) | 106(50.2) | | 106 (50.2) | 168 (47.3) | |
| Male vaccination status | | | 0.001 | | | 0.504 |
| no | 191 (90.5) | 340 (66.5) | | 315 (88.7) | 191 (90.5) | |
| yes | 20 (9.5) | 171 (33.5) | | 40 (11.3) | 20 (9.5) | |

The variables in PSM model included infertility duration、infertility diseases、ovarian stimulation protocols、male vaccination status and age.

Table 2 Embryo development and pregnancy outcomes by the female vaccination status

| | Before PSM | | | After PSM | | |
|---|---|---|---|---|---|---|
| | Vaccinated | Unvaccinated | P-value | Vaccinated | Unvaccinated | P-value |
| n | 511 | 211 | | 355 | 211 | |
| No. of ≥14 mm follicles on trigger day | 9(7,11) | 9(7,12) | 0.604 | 9 (7,12) | 9 (7,11) | 0.564 |
| No. of ≥17 mm follicles on trigger day | 5(4,7) | 5(4,7) | 0.559 | 5 (4,7) | 5 (4,7) | 0.688 |
| No. of follicles | 12.0 (9.0,15.0) | 12.0 (9.0,15.0) | 0.758 | 12.0 (9.0,15.0) | 12.0 (9.5,15.0) | 0.613 |
| No. of oocytes retrieved | 10(7,13) | 10(8,13) | 0.842 | 10 (7,13) | 10 (8,13) | 0.911 |
| No. of mature oocytes | 7.0 (3.0,11.0) | 8.0 (4.0,11.0) | 0.150 | 8.0 (4.0,11.0) | 8.0 (4.5,11.0) | 0.783 |
| No. of fertilized eggs | 7(5,10) | 8(5,11) | 0.706 | 10 (7,13) | 10 (8,13) | 0.911 |
| No. of MII oocytes | 10(7,13) | 10(7,13) | 0.587 | 10 (7,13) | 10 (7,13) | 0.931 |
| No. of 2PN oocytes | 7(4,9) | 7(4,9) | 0.669 | 6 (4,10) | 7 (4,9) | 0.588 |
| No. of cleaved embryos | 6(4,9) | 7(4,9) | 0.758 | 6 (4,9) | 7 (4,8.5) | 0.666 |
| No. of high-quality embryos on day 3 | 2(1,4) | 2(1,4) | 0.468 | 2 (1,4) | 2 (1,4) | 0.594 |
| No. of high-quality blastocysts | 1(0,4) | 2(0,4) | 0.801 | 2 (0,4) | 2 (0,4) | 0.897 |
| No. of blastocytes | 5(2,8) | 5(2,8) | 0.920 | 5 (2,8) | 5 (2,8) | 0.905 |
| Biochemical pregnancy | 351/511 (68.7) | 132/211 (62.6) | 0.111 | 246/355 (69.3) | 132/211 (62.6) | 0.100 |
| Clinical pregnancy | 302/511 (59.1) | 111/211 (52.6) | 0.109 | 212/355 (59.8) | 111/211 (52.6) | 0.098 |
| Ongoing pregnancy | 279/511 (54.6) | 105/211 (49.7) | 0.167 | 192/355 (54.1) | 105/211 (49.8) | 0.081 |

Table 3 Baseline characteristics stratified by male vaccination status.

| | Before PSM | | | After PSM | | |
|---|---|---|---|---|---|---|
| | Vaccinated | Unvaccinated | P-value | Vaccinated | Unvaccinated | P-value |
| n | 191 | 531 | | 190 | 360 | |
| BMI (kg/m2) | 22.1 (20.2,24.7) | 21.6 (20.0,23.9) | 0.070 | 22.1 (20.3,24.8) | 21.8 (20.0,24.2) | 0.175 |
| AMH | 2.3 (2.0,4.1) | 3.0 (2.0,4.3) | 0.343 | 2.9 (2,4.12.0) | 3.0 (2.1,4.5) | 0.106 |
| Infertility duration (year) | 3 (1,5) | 2 (1,4) | 0.081 | 3 (1,5) | 3 (2,4) | 0.374 |
| Infertility diseases, n (%) | | | 0.369 | | | 0.231 |
| Tubal factor | 114 (59.7) | 303 (57.1) | | 113 (59.5) | 200 (55.6) | |
| Ovarian factor | 16 (8.4) | 34 (6.4) | | 16 (8.4) | 19 (5.3) | |
| Male factor | 34 (17.8) | 91 (17.1) | | 34 (17.9) | 72 (2.0) | |
| Unexplained | 27 (14.1) | 103 (19.4) | | 27 (14.2) | 69 (19.2) | |
| Type of infertility, n(%) | | | 0.343 | | | 0.396 |
| Primary | 92 (48.2) | 277 (52.2) | | 92 (48.4) | 188 (52.2) | |
| Secondary | 99 (51.8) | 254 (47.8) | | 98 (51.6) | 172 (47.8) | |
| Ovarian stimulation protocols, n(%) | | | 0.321 | | | 0.868 |
| Antagonist | 99 (51.8) | 253 (47.7) | | 92 (48.4) | 177 (49.2) | |
| Agonist | 92 (48.2) | 278 (52.4) | | 98 (51.6) | 183 (50.8) | |
| Age (years) | 31 (28,34) | 31 (28,34) | 0.578 | 31 (28,34) | 31 (28,34) | 0.539 |
| Gravidity | 0 (0,2) | 1 (0,2) | 0.259 | 0 (0,2) | 1 (0,2) | 0.983 |
| Parity | 0 (0,0) | 0 (0,0) | 0.857 | 0 (0,2) | 1 (0,2) | 0.264 |
| Basal FSH | 7.7 (6.5,8.7) | 7.4 (6.3,8.5) | 0.103 | 7.7 (6.5,8.7) | 7.4 (6.4,8.9) | 0.145 |
| Basal E2 | 32.0 (25.0,41.0) | 33.0 (26.0,41.0) | 0.358 | 3.02 (25.3,41.0) | 33.0 (26.0,41.0) | 0.531 |
| Basal P | 0.5 (0.4,0.7) | 0.5 (0.3,0.8) | 0.686 | 0.5 (0.4,0.7) | 0.5 (0.3,0.8) | 0.919 |
| Basal PRL | 256.9 (194.1,342.7) | 258.0 (180.3,359.4) | 0.872 | 256.4 (193.8,341.8) | 252.4 (177.2,356.1) | 0.759 |
| Basal LH | 3.5 (2.7,4.8) | 3.7 (2.6,4.9) | 0.701 | 3.5 (2.7,4.8) | 3.6 (2.7,4.9) | 0.792 |
| Total AFC | 15 (10,21) | 15 (11,20) | 0.314 | 15 (10,21) | 15 (11,21) | 0.284 |
| LH level on trigger day (mIU/mL) | 1.1 (0.7,2.4) | 1.1(0.7,2.1) | 0.785 | 1.0 (0.7,2.3) | 1.2 (0.7,2.2) | 0.848 |
| P level on trigger day (ng/mL) | 0.6(0.4,0.9) | 0.7(0.5,1.0) | 0.033 | 0.6 (0.4,0.9) | 0.7 (0.4,0.9) | 0.149 |
| E2 level on trigger day (pg/mL) | 1667.0(1271.0,2413.0) | 1847.0(1350.0,260.01) | 0.071 | 1653.5(1269.8,2394.8) | 1863.0(1357.0,2615.3) | 0.044 |
| Fertilization type, n(%) | | | 0.056 | | | 0.287 |
| IVF | 145 (75.9) | 437 (82.3) | | 144 (75.8) | 287 (79.7) | |
| ICSI | 46 (24.1) | 94 (17.7) | | 46 (24.2) | 73 (20.3) | |
| Endometrial thickness(mm) on the | 11.0(9.5,12.5) | 11.0(9.5,12.0) | 0.829 | 11.0 (9.5,12.5) | 11.0 (9.5,12.0) | 0.994 |
| Total gonadotropin dose (IU) | 2025 (1650,2475) | 2025 (1650,2419) | 0.923 | 2025 (1650,2475) | 2025 (1650,2400) | 0.654 |
| Stimulation duration (days) | 10 (9,12) | 10 (9,12) | 0.452 | 10 (9,12) | 10 (9,12) | 0.804 |
| No. of embryos transferred, n (%) | | | 0.912 | | | 0.716 |
| 1 | 70 (36.7) | 197 (37.1) | | 70 (36.8) | 127 (35.3) | |
| 2 | 121 (63.4) | 334 (62.9) | | 120 (63.2) | 233 (64.7) | |
| No. of high-quality embryos for transfer, n (%) | | | 0.364 | | | 0.549 |
| 0 | 42 (22.0) | 92 (17.3) | | 41 (21.6) | 64 (17.8) | |
| 1 | 95 (49.7) | 279 (52.5) | | 95 (50.0) | 186 (51.7) | |
| 2 | 54 (28.3) | 160 (30.1) | | 54 (28.4) | 110 (30.6) | |
| Days of embryos transferred, n (%) | | | 0.691 | | | 0.535 |
| D3 | 95 (49.7) | 273 (51.4) | | 95 (50.0) | 190 (52.8) | |
| D5 | 96 (50.3) | 258 (48.6) | | 95 (50.0) | 170 (47.2) | |
| Female vaccination status | | | 0.001 | | | 0.834 |
| no | 20 (10.5) | 191 (36.0) | | 20 (10.5) | 40 (11.1) | |
| yes | 171 (89.5) | 340 (64.0) | | 170 (89.5) | 320 (88.9) | |

The variables in PSM model included P level on trigger day, age, female vaccination status and fertilization type.

Table 4 Embryo development and pregnancy outcomes stratified by male vaccination status.

| | Before PSM | | | After PSM | | |
|---|---|---|---|---|---|---|
| | Vaccinated | Unvaccinated | P-value | Vaccinated | Unvaccinated | P-value |
| n | 191 | 531 | | 190 | 360 | |
| No. of ≥14 mm follicles on trigger day | 9.0 (6.0,12.0) | 9.0 (7.0,11.0) | 0.397 | 8.5(6.0,11.8) | 9.0(7.0,11.23) | 0.264 |
| No. of ≥17 mm follicles on trigger day | 5 (4,7) | 5 (4,7) | 0.294 | 5 (4,7) | 5 (4,7) | 0.317 |
| No. of follicles | 12 (9,15) | 12 (9,15) | 0.297 | 11.5 (9,15) | 12 (9,15) | 0.225 |
| No. of oocytes retrieved | 9.0 (7.0,12.0) | 10.0 (8.0,13.0) | 0.012 | 9 (6.3,12.0) | 11.0(7.0,14.0) | 0.013 |
| No. of mature oocytes | 7 (2,10) | 8 (4,11) | 0.057 | 7 (2,10) | 8 (3,11) | 0.219 |
| No. of high-quality embryos on day 3 | 2 (1,4) | 2 (1,4) | 0.219 | 2 (1,4) | 2 (1,4) | 0.640 |
| No. of high-quality blastocysts | 1.0 (0.0,4.0) | 2.0 (0.0,4.0) | 0.106 | 1.0 (0.0,3.8) | 1.0 (0.0,4.0) | 0.230 |
| No. of blastocytes | 5 (2,8) | 5 (2,8) | 0.161 | 5 (2,8) | 5 (2,8) | 0.219 |
| Biochemical pregnancy | 126/191(66.0) | 357/531 (67.2) | 0.750 | 126/190(66.3) | 244/360(67.8) | 0.728 |
| Clinical pregnancy | 108/191(56.5) | 305/531 (57.4) | 0.830 | 108/190(56.8) | 211/360(58.6) | 0.689 |
| Ongoing pregnancy | 105/191(55.0) | 279/531 (52.5) | 0.663 | 105/190 (55.3) | 189/360(52.5) | 0.760 |

Table 5 Baseline characteristics of couples both vaccinated against COVID-19

| | Vaccinated | Unvaccinated | P-value |
|---|---|---|---|
| n | 171 | 191 | |
| BMI (kg/m2) | 21.5 (20.0,23.4) | 22.2 (20.3,24.8) | 0.036 |
| AMH (ng/ml) | 2.8 (20.0,4.) | 2.93 (2.1,4.2) | 0.622 |
| Infertility duration (years) | 2.0 (1.0,4.0) | 3.0 (1.5,5.0) | 0.021 |
| Infertility diseases, n (%) | | | 0.055 |
| Tubal factor | 115 (60.2) | 102 (59.7) | |
| Ovarian factor | 14 (7.3) | 16 (9.4) | |
| Male factor | 22 (11.52 | 32 (18.7) | |
| Unexplained | 40 (20.9) | 21 (12.3) | |
| Type of infertility, n (%)) | | | 0.409 |
| Primary | 101 (52.9) | 83 (48.5) | |
| Secondary | 90 (47.1) | 88 (51.5) | |
| Ovarian stimulation protocols, n (%) | | | 0.039 |
| Agonist | 109 (57.1) | 79 (46.2) | |
| Antagonist | 82 (42.9) | 92 (53.8) | |
| Age (years) | 31 (28,3) | 31 (28,3) | 0.561 |
| Parity | 1 (0,2) | 0 (0,2) | 0.211 |
| Gravidity | 0 (0,0) | 0 (0,0) | 0.896 |
| Basal FSH | 7.2 (6.3,8.5) | 7.7 (6.4,8.6) | 0.335 |
| Basal E2 | 32 (26,40) | 32 (25,40) | 0.581 |
| Basal P | 0.5 (0.3,0.7) | 0.5 (0.4,0.7) | 0.801 |
| Basal PRL | 269.7 (185.3,371.3) | 257.4 (193.3,344.0) | 0.824 |
| Basal LH | 3.7 (2.5,5.2) | 3.5 (2.6,4.8) | 0.763 |
| Total AFC | 15.0 (11.0,19.0) | 15.0 (9.5,20.0) | 0.708 |
| LH level on trigger day (mIU/mL) | 1.1 (0.7,1.8) | 1.04 (0.7,2.4) | 0.517 |
| P level on trigger day (ng/mL) | 0.7 (0.4,0.9) | 0.6 (0.4,0.9) | 0.252 |
| E2 level on trigger day (pg/mL) | 1842.0 (1321.0,2586.0) | 1640.0 (1275.5,2413.0) | 0.264 |
| Fertilization type, n (%)) | | | 0.024 |
| ICSI | 27 (14.1) | 40 (23.4) | |
| IVF | 164 (85.9) | 131 (76.6) | |
| No. of ≥14 mm follicles on trigger day | 9.0 (7.0,11.0) | 9.0 (6.5,12.0) | 0.859 |
| No. of ≥17 mm follicles on trigger day | 5 (4,7) | 5 (4,7) | 0.306 |
| Endometrial thickness(mm) on the | 11.0 (9.8,12.5) | 10.5 (9.5,12.0) | 0.344 |
| Total gonadotropin dose (IU) | 2025 (1650,2475) | 1950 (1650,2475) | 0.597 |
| Stimulation duration (days) | 10 (9,12) | 10 (9,12) | 0.103 |
| No. of embryos transferred, n (%) | | | 0.654 |

|   | 1 | 77 (40.3) | 65 (38.0) |   |
|---|---|---|---|---|
|   | 2 | 114 (59.7) | 106 (62.0) |   |
| Days of embryos transferred, n (%) |   |   |   | 0.949 |
|   | D3 | 90 (47.1) | 80 (46.8) |   |
|   | D5 | 101 (52.9) | 91 (53.2) |   |

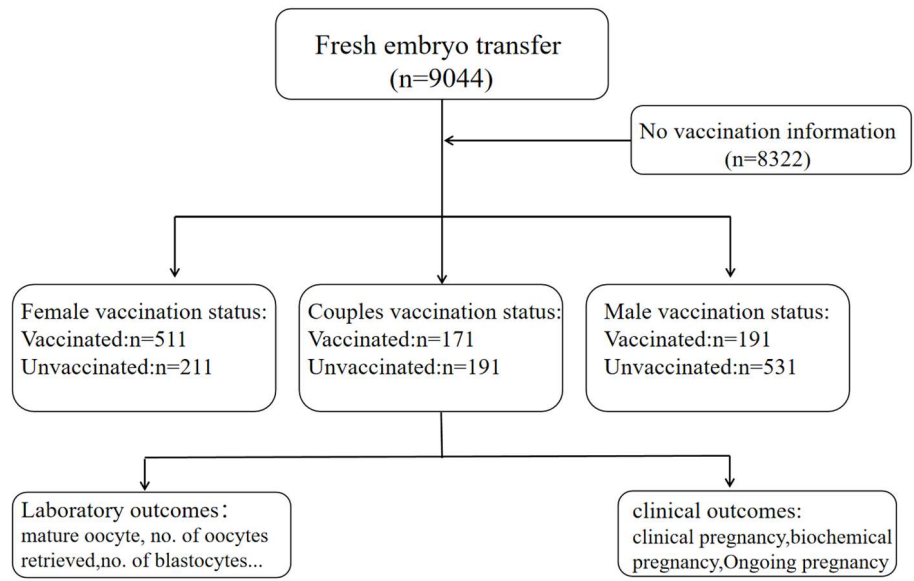

Figure 1 flowchart of the study

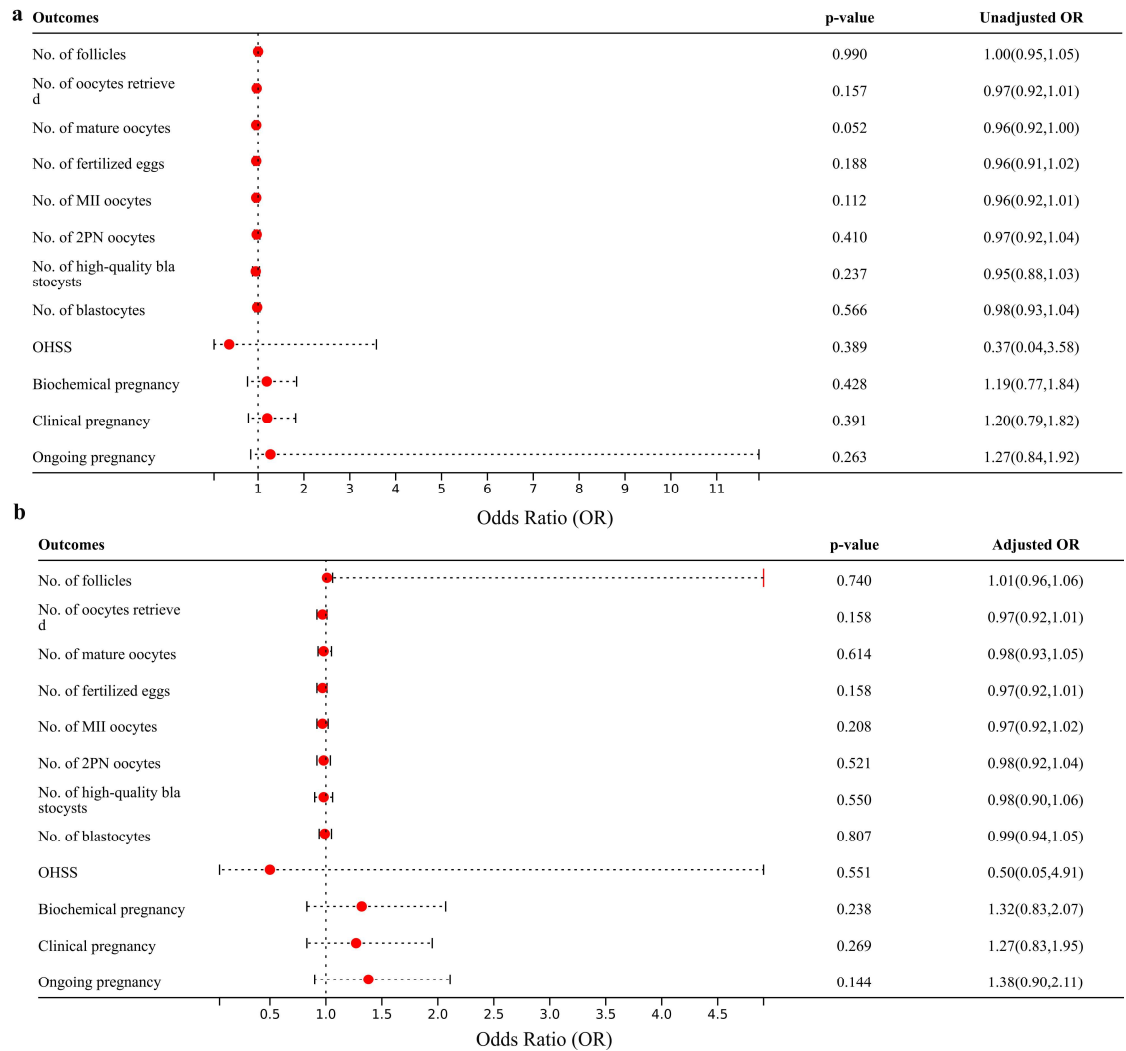

Figure2 Logistic regression analysis of pregnancy outcomes in vaccinated couples.

a: Unadjusted OR. b: Adjusted OR. Adjusted for BMI, infertility duration, ovarian stimulation protocols, fertilization type, age. OR, odds ratio.

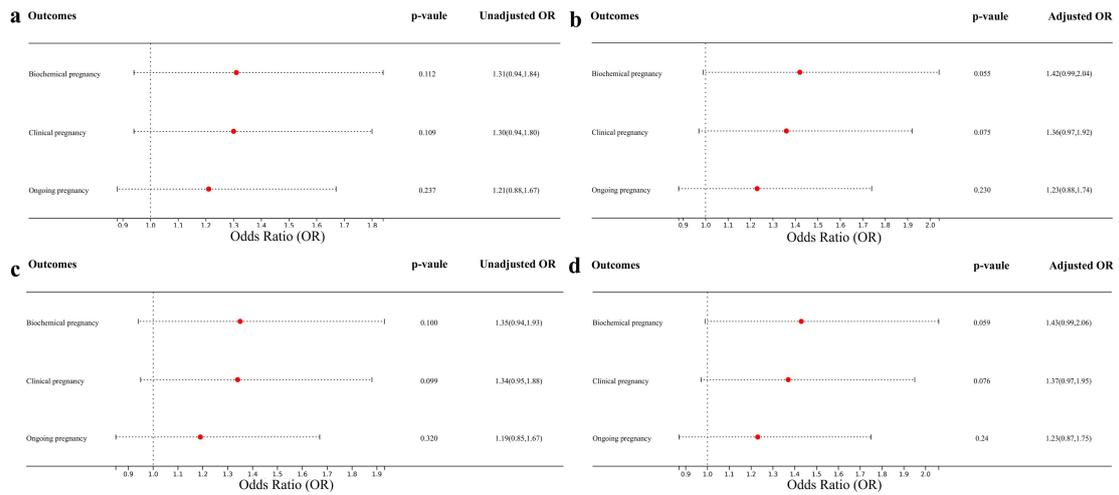

Figure 3 Logistic regression compares pregnancy outcomes in vaccinated and unvaccinated females, pre- and post-PSM.

a: Unadjusted Odds Ratio before PSM. b: Adjusted OR before PSM. c: Unadjusted Odds Ratio after PSM. d: Adjusted OR after PSM. OR, odds ratio. b and d adjusted for male vaccination status, infertility duration, infertility diseases, AMH and ovarian stimulation protocols.

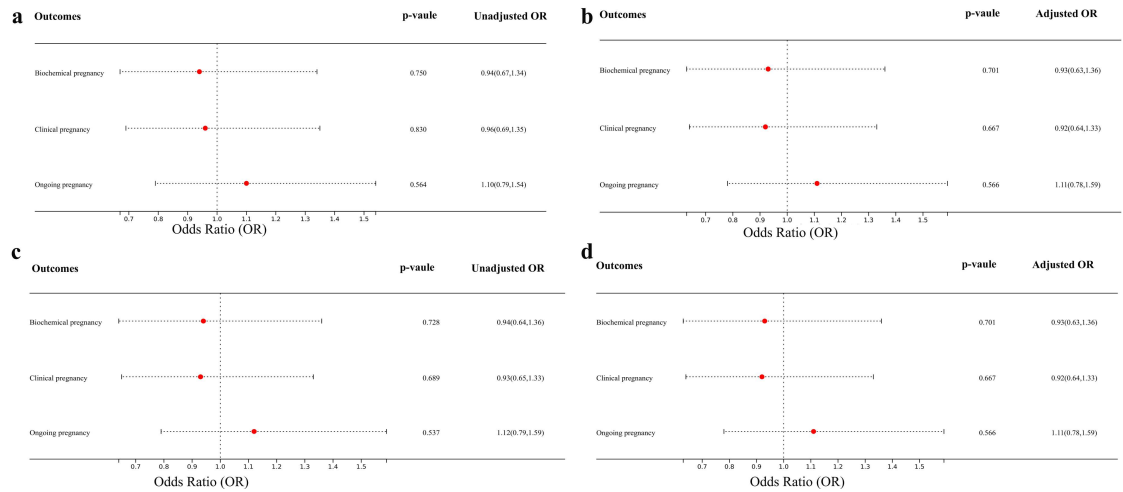

Figure4 Logistic regression compares pregnancy outcomes in vaccinated and unvaccinated males, pre- and post-PSM.

a: Unadjusted Odds Ratio before PSM. b: Adjusted OR before PSM. c: Unadjusted Odds Ratio after PSM. d: Adjusted OR after PSM. b and d Adjusted for female vaccination status, age, P level on trigger day, E2 level on trigger day and fertilization type. OR, odds ratio.